**Towards accurate predictions of bond-selective fluorescence spectra**

Philip A. Kocheril, Ryan E. Leighton, Noor Naji, Dongkwan Lee, Haomin Wang, Jiajun Du, and Lu Wei*

Division of Chemistry and Chemical Engineering, California Institute of Technology, Pasadena, CA 91125, USA

*Corresponding author: lwei@caltech.edu

**ABSTRACT**

Vibrational-encoded fluorescence spectro-microscopies are emerging as powerful tools for studying molecular vibrations with the unparalleled sensitivity of fluorescence spectroscopy. We recently described one such technique, termed bond-selective fluorescence-detected infrared-excited (BonFIRE) spectro-microscopy. Currently, prospects of BonFIRE towards rational molecular design are limited, but they have the potential to be assisted by computational tools. In this Perspective, we provide a brief overview of the theory of BonFIRE spectroscopy. We then describe a fully automated computational pipeline for calculating BonFIRE spectra, reproducing key features of experimental results. Finally, we highlight a few potential applications of computational methods for vibrational-encoded fluorescence spectro-microscopies and their broader implications for chemistry and biology.

# I. INTRODUCTION

Molecular vibrations are established, quantitative probes of rich chemical information,[1] but traditional vibrational spectroscopies – most commonly, infrared (IR) absorption and Raman scattering – are inherently limited in sensitivity. In the 1970s, Kaiser and co-workers pioneered a vibrational-electronic double-resonance fluorescence technique,[2] allowing molecular vibrations to be detected through changes in visible fluorescence and yielding much improved sensitivity. This technique, alongside others, granted valuable insights into the nature and dynamics of molecular vibrations[3, 4] and served as a basis for early super-resolution IR imaging.[5, 6] Over the past decade, fluorescence-detected vibrational spectroscopies have seen much renewed interest,[7, 8] demonstrating real-world applications from biological imaging to materials science.[9-12]

We recently described one such method, termed bond-selective fluorescence-detected IR-excited (BonFIRE) spectro-microscopy (**Fig. 1**).[13] BonFIRE is a two-pulse experiment, where a mid-IR (MIR) pulse excites a vibrational population and a near-IR (NIR) pulse up-converts the vibrationally excited population to a fluorescently active one (**Fig. 1a**). The bond-selectivity of BonFIRE results from using narrowband (~10 cm$^{-1}$), 2-ps pulses, leading to efficient excitation of a specific vibrational mode. BonFIRE is an action spectroscopy, where changes in the absorption of the MIR and NIR pulses are inferred through changes in integrated fluorescent intensity (under the assumption that the quantum yield and emission profile are constant, which has been validated elsewhere[10]).

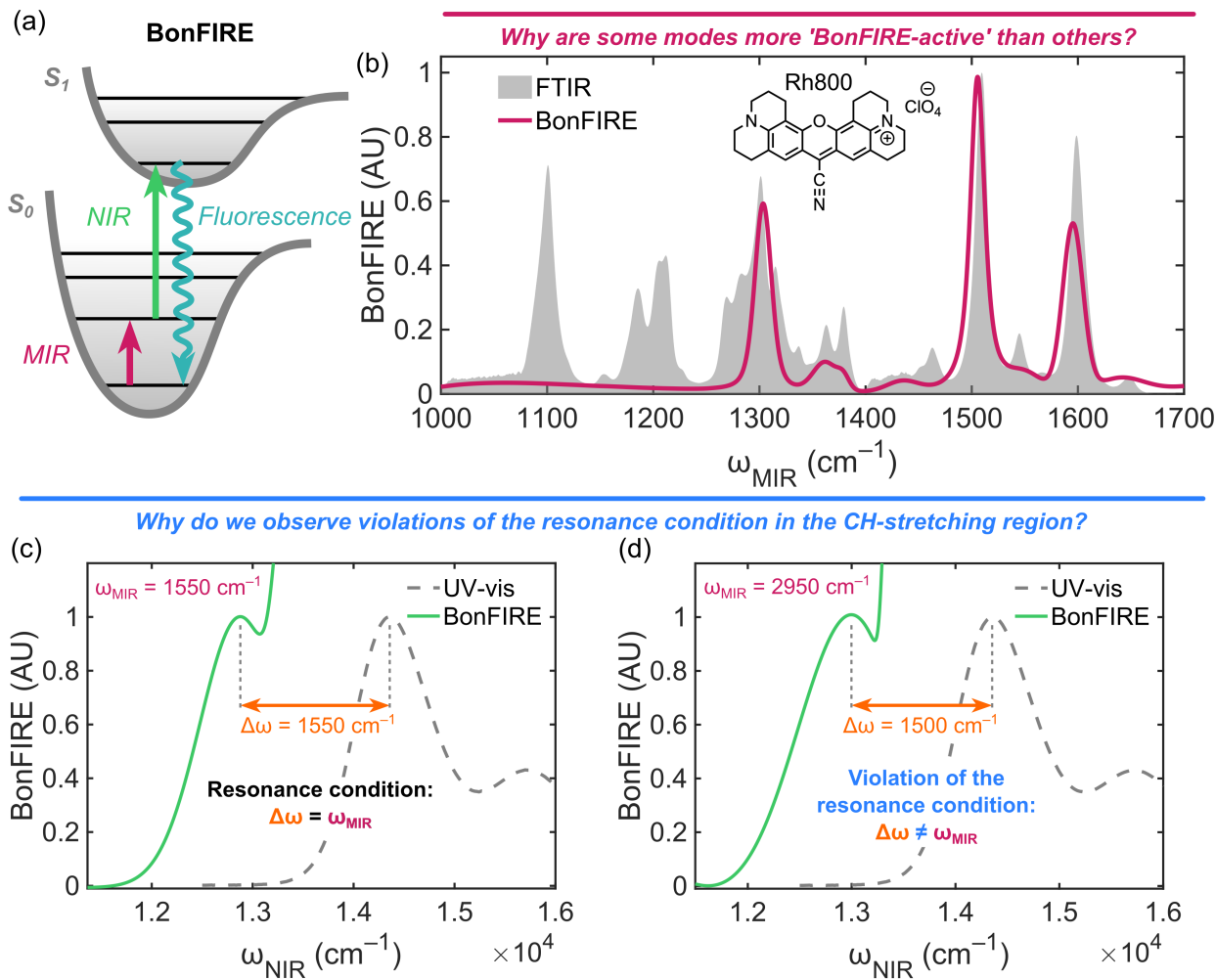

**Figure 1.** Overview of BonFIRE spectroscopy and open questions to be addressed with computational tools. (a) Principle of BonFIRE, comprising narrowband, mode-selective MIR and NIR excitations. (b) MIR frequency-dependence in BonFIRE for Rh800 (structure inset).[14] (c) NIR frequency-dependence in BonFIRE in the fingerprint region for Rh800, demonstrating the resonance condition.[14] (d) NIR frequency-dependence in the CH-stretching region for Rh800, exhibiting a violation of the resonance condition (VRC).[14] All spectra are normalized.

Thus far, our work with BonFIRE has been primarily experiment-driven, towards functional spectroscopy and bioimaging with single-molecule sensitivity.[13-17] As such, we believe there is significant opportunity for computational methods (whether quantum, classical, or machine-learned) to aid with interpretation of experimental spectra and act as a guide towards future experiments. Importantly, such tools[18] have only recently begun to be used for vibrational-electronic double-resonance spectra but have demonstrated reasonable quantitative accuracy for coumarin dyes and other small, blue-absorbing chromophores.[19-22]

There are several open questions in BonFIRE where computational tools can provide valuable insights. For example, why are some vibrational modes more 'BonFIRE-active' than others? This is readily apparent when examining the MIR frequency ($\omega_{MIR}$)-dependence in BonFIRE (**Fig. 1b**) for a model dye, Rhodamine 800 (Rh800), which resembles a Fourier transform IR (FTIR) absorption spectrum.[13] The differences in relative intensity between different peaks are generally attributed to the vibronic coupling strengths of the different vibrational modes;[23] however, it is highly nontrivial to assess if a vibrational mode will be strongly coupled by examining molecular structure, with π-conjugation and delocalized normal mode atomic displacements not being reliable indicators, as we and others have shown previously.[19, 24]

Another open question emerges from examining the NIR frequency ($\omega_{NIR}$)-dependence in BonFIRE (**Fig. 1c-d**). For most vibrational modes, the $\omega_{NIR}$ spectrum resembles an ultraviolet-visible (UV-vis) absorbance spectrum in shape, but its peak is redshifted by the energy of the absorbed MIR photon (**Fig. 1c**), which has been termed the "resonance condition" in previous vibrational-encoded fluorescence work.[25] However, in our most recent report, we identified several BonFIRE-active modes in the CH-stretching region (2600-3200 cm$^{-1}$) that instead exhibited a redshift of a much smaller energy than the absorbed MIR photon (**Fig. 1d**; termed a "violation of the resonance condition," VRC).[14] From a series of experimental controls, we reasoned that these features should originate from MIR combination modes rather than CH-stretches, but the mechanistic basis for VRCs remained inconclusive.[14]

Furthermore, several aspects of molecular probe design with BonFIRE remain largely unexplored. For example, we have recently demonstrated single-shot 16-color imaging with BonFIRE[14] and shown that nitrile vibrational lifetimes can act as reporters of local electric fields.[15] To advance further in each of these aims, we see significant potential for computations to guide in designing molecular probes with well-separated BonFIRE peaks (towards super-multiplex imaging with more colors) and to fuel discovery of new, higher-performance local sensors.

In this Perspective, we explore methods for computing BonFIRE spectra from first principles to gain greater insight into our experimental results and discuss their predictive utility towards future molecular design efforts. We first provide a brief description of the underlying theoretical principles for calculating vibrationally resolved electronic spectra. We then show that density functional theory (DFT) calculations can accurately reproduce key features of both steady-state (FTIR, UV-vis, and fluorescence) and BonFIRE spectra. To facilitate high-throughput *in silico* screening, we developed a fully automated pipeline for BonFIRE spectrum predictions, integrating several industry-standard computational chemistry tools (OpenBabel, Gaussian16, and FCclasses3)[18, 26, 27] into an openly accessible Python wrapper and requiring only an input ChemDraw structure for all calculations. By computing electronic absorption spectra following vibrational pre-excitation, we demonstrate that these computations provide new insight into the underlying mechanisms of VRCs. Finally, we discuss several potential applications of vibronic spectrum calculations towards future work with BonFIRE.

## II. THEORY

In this section, we provide a brief overview of the theoretical basis for BonFIRE. The foundation of this section is a series of detailed papers from Tokmakoff and co-workers, who described several key aspects of the theory of vibrational-electronic double-resonance fluorescence spectroscopies.[19, 28, 29] Their work in turn builds on the foundational work of Santoro and co-workers on efficient methods for calculating vibrationally resolved electronic spectra within the harmonic approximation.[30-33]

Because we use narrowband pulses in BonFIRE, we make the simplifying assumption that multimode coherences are negligible,[19] indicating that we can treat spectra without response function theory.[28] We also neglect polarization effects and the impact of vibrational lifetimes here, adopting a steady-state picture.[29, 34] Thus, our goal is to calculate two quantities:

(1) MIR absorption cross-sections, and

(2) NIR absorption cross-sections following MIR excitation.

In addition to working with larger NIR dyes of greater structural diversity (rhodamine, oxazine, and cyanine scaffolds), our work expands upon previous work by broadening the scope of treated vibrational modes (here, 1,000-3200 cm$^{-1}$). Furthermore, we directly compare experimental BonFIRE $\omega_{NIR}$ spectra to computed $\omega_{NIR}$ spectra (obtained by the sum of many individual vibronic transitions), allowing us to identify and assign the major vibronic transitions contributing to signal in BonFIRE.

### A. BonFIRE as a double-resonance process

We first consider MIR excitation of a single vibrational mode ($|S_0^0\rangle \to |S_0^v\rangle$) and NIR excitation to the ground vibrational state of the electronic excited state ($|S_0^v\rangle \to |S_1^{0'}\rangle$), with

vibrational states in $S_1$ denoted with primes ($'$). We describe BonFIRE as two sequential, linear absorptions,[13] followed by fluorescence according to the molecule's fluorescence quantum yield ($\Phi$). Thus, for electric dipole-allowed transitions, BonFIRE intensity ($I_{BF}$) is proportional to the product of two (absolute-squared) transition dipole moments:[35]

$$I_{BF} \propto \Phi \cdot |\langle S_1^{0'}|\mu_e|S_0^v\rangle|^2 \cdot |\langle S_0^v|\mu_v|S_0^0\rangle|^2 \quad (1)$$

In **Eq. 1**, the right overlap integral describes the MIR absorption (mediated by vibrational dipole $\mu_v$) from $|S_0^0\rangle$ to $|S_0^v\rangle$, and the left overlap integral describes the NIR absorption (mediated by electronic dipole $\mu_e$) from $|S_0^v\rangle$ to $|S_1^{0'}\rangle$. The remaining constants of proportionality in **Eq. 1** (not written) include the concentration of the molecule, the fluorescence collection efficiency of the optics used (e.g., objectives, bandpass filters, dichroic beamsplitters), the detector's quantum efficiency, the optical powers, and more.[25]

Taking the MIR absorption as an example, the vibrational transition dipole moment is related to the MIR extinction coefficient ($\varepsilon_{MIR}(\omega)$):[36]

$$|\langle S_0^v|\mu_v|S_0^0\rangle|^2 = \frac{3\varepsilon_0 hc \ln 10}{2\pi^2 N_A} \int \frac{\varepsilon_{MIR}(\omega)}{\omega} d\omega \quad (2)$$

In **Eq. 2**, $\varepsilon_0$, $h$, $c$, $N_A$, and $\omega$ are the vacuum permittivity, Planck's constant, the speed of light, Avogadro's number, and the frequency in cm$^{-1}$, respectively.[36] Thus, the $\omega_{MIR}$-dependence in BonFIRE is directly related to the MIR extinction coefficient, in good agreement with what we observe experimentally (**Fig. 1b**).

### B. Electronic absorption following vibrational excitation

Originating from Franck's mechanism for photodissociation[37] and Condon's application to intensity distributions,[38, 39] the problem of vibrationally resolved electronic spectra has seen much interest over the past century.[40-42] Applying the Born-Oppenheimer approximation, the electronic transition dipole moment in **Eq. 1** can be simplified by separating the vibronic wavefunctions into products of electronic and vibrational wavefunctions ($|S_0^v\rangle = |S_0\rangle|v\rangle$).[19] Within the Condon approximation, the electronic transition dipole ($\mu_e$) is independent of the vibrational coordinates,[24] meaning that we can write:

$$\langle S_1^{0'}|\mu_e|S_0^v\rangle = \langle S_1|\mu_e|S_0\rangle \cdot \langle 0'|v\rangle \quad (3)$$

**Eq. 3** shows that a vibronic transition dipole moment can be factored into a purely electronic transition dipole moment ($\langle S_1|\mu_e|S_0\rangle$) and a purely vibrational component describing the overlap of the initial and final vibrational wavefunctions ($\langle 0'|v\rangle$), the latter of which is known as the Franck-Condon factor (FCF). Conceptually, the electronic transition dipole moment determines the overall spectral intensity, and the FCFs determine the shape of the spectrum (i.e., the intensities of individual transitions relative to others in the same band). FCFs may be positive or negative, depending on the phase relationship of the vibrational wavefunctions, but the absolute-squared FCF (ranging from 0 to 1) is proportional to the transition intensity.

For the electronic absorption, we can define the up-conversion cross-section relative to the extinction coefficient of the 0-0 transition ($\varepsilon_{00} = |\langle S_1|\mu_e|S_0\rangle|^2|\langle 0'|0\rangle|^2$), which allows us to substitute out the electronic transition dipole moment. Thus, BonFIRE intensity for the double-resonance process $|S_0^0\rangle \to |S_0^v\rangle \to |S_1^{0'}\rangle$ can be rewritten as:

$$I_{BF} \propto \Phi \cdot \varepsilon_{00} \cdot \left[\frac{|\langle 0'|v\rangle|^2}{|\langle 0'|0\rangle|^2}\right] \cdot \varepsilon_{MIR} \quad (4)$$

The term in square brackets in **Eq. 4** is sometimes called the Franck-Condon ratio (FCR),[23] describing the vibronic coupling strength of a specific transition relative to that of the 0-0 transition. The FCR is a particularly desirable quantity for comparing vibronic intensities between different molecules, since the remaining terms in **Eq. 4** can all be obtained experimentally through steady-state measurements (FTIR, UV-vis, and fluorescence spectroscopies).

## C. Calculation of spectra

To extend our treatment from **Eq. 4** (for a single vibrational mode and up-conversion only to $|S_1^{0'}\rangle$) to a general form (for all modes), we adopt a time-independent formalism, summing over all possible MIR-excited states in $S_0$ ($i$; $3N-6$ normal modes for an $N$-atom nonlinear molecule) and all possible final states in $S_1$ ($j'$). We treat the resonance conditions as Dirac delta functions, with $\delta_{MIR} = \delta(\omega_i - \omega_{MIR})$ and $\delta_{NIR} = \delta(\omega_{j'} + \omega_{00} - \omega_{NIR} - \omega_{MIR})$. We thus arrive at a general expression for BonFIRE spectral intensity:

$$I_{BF} \propto \frac{\Phi \cdot \varepsilon_{00}}{|\langle 0'|0\rangle|^2} \cdot \sum_{j=0}^{3N-6} \sum_{i=1}^{3N-6} |\langle j'|i\rangle|^2 \cdot \varepsilon_{MIR}(\omega_i) \cdot \delta_{MIR}(i) \cdot \delta_{NIR}(i,j') \tag{5}$$

To calculate spectra, the task at hand then becomes computing each of the quantities in **Eq. 5** for each vibrational mode of interest. The resonance conditions can be subsequently broadened with Gaussian lineshapes to more closely resemble experimental spectra. For BonFIRE spectra of a single molecular species, $\Phi$, $\varepsilon_{00}$, $|\langle 0'|0\rangle|^2$ and the remaining constants of proportionality are taken to be the same. Thus, the frequency-dependences of BonFIRE are governed by the MIR absorption cross-sections and the FCFs.

## III. METHODS

### A. Computational methods

Detailed computational methods are given in **Section S1.1** of the Supplementary Material and summarized here.[15, 19, 26, 43] We use density functional theory (DFT) and time-dependent DFT (TDDFT) to calculate ground-state and excited-state electronic structure, as implemented in Gaussian 16 (rev. B.01).[27] The outputs of our DFT and TDDFT calculations are used to prepare input files for Cerezo and Santoro's FCclasses3 (v3.0.3).[18] Because we are interested in the FCFs of all of the vibrational modes of a molecule, we adopt a time-independent, sum-over-states formalism for computing electronic spectra in FCclasses3.[24] To generate BonFIRE spectra, we calculate electronic one-photon absorption with vibrational pre-excitation in each mode with FCclasses3, an extension of the method described for vibrationally promoted electronic resonance spectroscopy.[20, 44]

### B. Experimental methods

In this work, we analyze previously reported BonFIRE, FTIR, UV-vis absorption, and fluorescence emission spectra.[13-17, 45] Methods for these experiments are detailed in **Section S1.2** and briefly summarized here. Steady-state measurements were performed with 10 μM (for UV-vis and fluorescence; Varian Cary 500, Agilent and RF-6000, Shimadzu) and 100 mM (for FTIR; Vertex 80v, Bruker) solutions in dimethyl sulfoxide (DMSO; 276855, Millipore Sigma).[15, 17] For BonFIRE experiments, Rh800 (83701, Millipore Sigma), sulfo-cyanine5.5 (Cy5.5; A7330, Lumiprobe) and ATTO680 (AD 680, ATTO-Tec) were dissolved at 100 μM in DMSO-$d_6$ (151874, Cambridge Isotope Laboratories) or DMSO and sandwiched between two thin $CaF_2$ windows (CAFP10-0.35 and CAFP25-0.5, Crystran) separated by a 6-µm Teflon spacer (MSP-006-M13, Harrick Scientific).[14]

BonFIRE experiments (**Fig. S1**) were carried out with two independently tunable, synchronously pumped optical parametric oscillators (Levante IR and picoEmerald S, Applied Physics and Electronics) and a difference frequency generation unit (HarmoniXX DFG, Applied Physics and Electronics) pumped by a 1.6-ps, 1031.2-nm, 80-MHz, Yb fiber laser (aeroPULSE PS10, NKT Photonics).[13-17] The MIR pulse trains were modulated by acousto-optic modulators (GEM-40-4-4500/4mm and GEM-40-4-6282/2mm, Brimrose) and focused onto the sample with a ZnSe lens (39-469, Edmund Optics). The NIR pulse train was reflected off a retroreflector on a delay stage (DL-BKIT2U-S-M, Newport) to control the relative timing of the MIR and NIR pulses and focused onto the sample with a water-immersion 25× objective lens (XLPLN25XWMP2, Olympus). Fluorescence emission was spectrally filtered (FF738-Di01 and FF01-665/150, Semrock) and focused (AC254-200-B, Thorlabs) onto a photomultiplier tube (PMT1002, Thorlabs), the output of which was sent to a fast lock-in amplifier (HF2LI, Zurich Instruments) and demodulated with a reference signal (time constant 3 ms). Data acquisition was controlled by a custom LabVIEW virtual instrument (v19.0f2; National Instruments).

## IV. RESULTS AND DISCUSSION

### A. Steady-state spectra

We must first ensure that our computational methods yield accurate and robust calculations of vibrational and electronic structure. Consistent with our previous work, we find that DFT calculations with the Becke three-parameter, Lee-Yang-Parr (B3LYP) functional[46] and 6-31G(d,p) basis set provide a good description of ground-state vibrational structure in large polyatomic dyes like Rh800, as compared to an experimental FTIR spectrum in DMSO (**Fig. 2a**).[15] The agreement in normal mode frequencies between theory and experiment is quite reasonable (within ~5 cm$^{-1}$ after scaling), and although the calculated intensities agree less well with experimental spectra (within an order of magnitude), the relative sorting of strong vs. weak modes is reproduced.

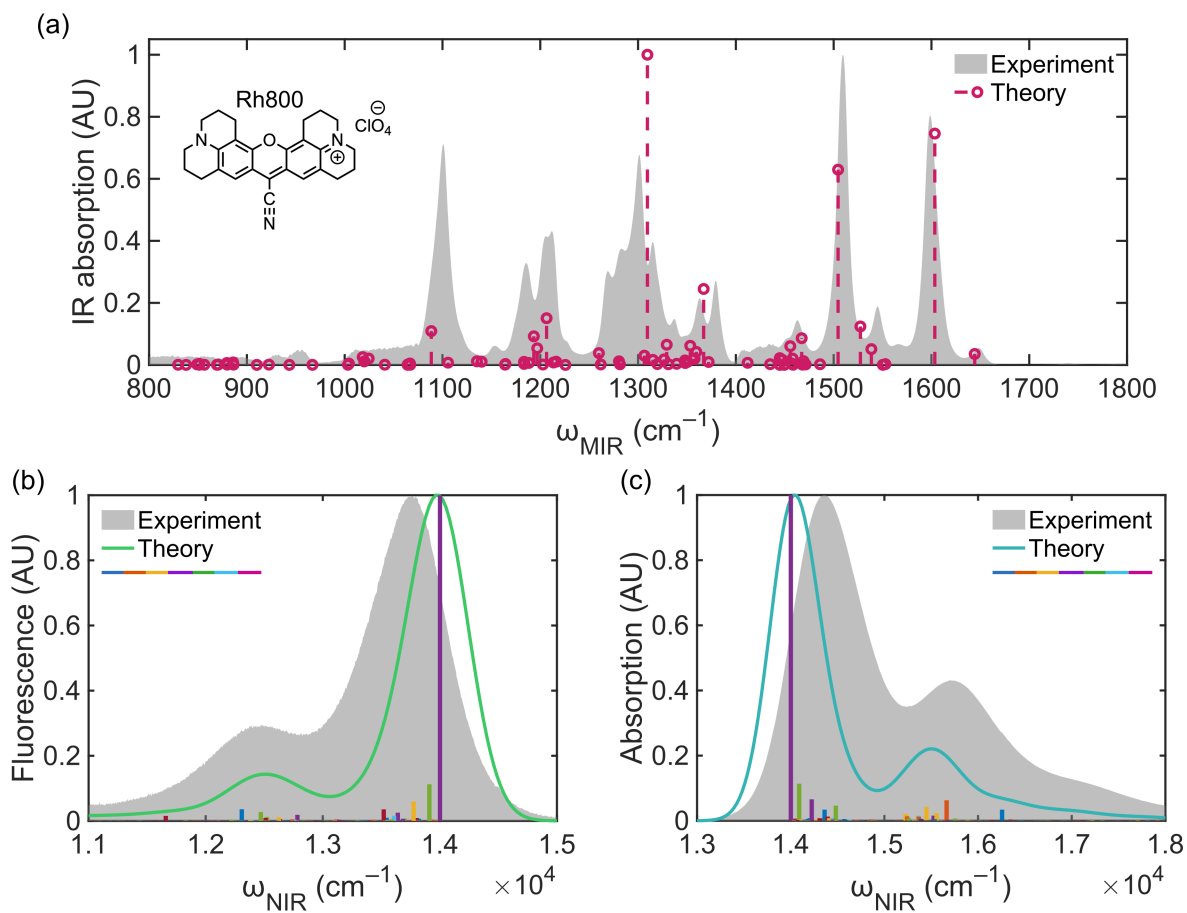

**Figure 2.** Calculations of steady-state spectra of Rh800. (a) Comparison of predicted IR absorption at the B3LYP/6-31G(d,p) level (frequencies scaled by 0.97) against an experimental FTIR spectrum (100 mM Rh800 in DMSO; structure inset). (b-c) Comparison of predicted (b) fluorescence emission and (c) electronic one-photon absorption spectra against experimental fluorescence and UV-vis spectra (10 μM Rh800 in DMSO). Spectra were calculated from structures optimized at the B3LYP/6-31G(d,p) level with the SMD solvent model, using FCclasses3. We plot stick spectra as colored vertical lines to show the individual vibronic transitions (frequencies unscaled), with the colored curves showing the combined spectrum with 290 cm$^{-1}$ (HWHM) of Gaussian broadening. Experimental spectra were obtained from a previous report.[15] All spectra are normalized.

We then calculated fluorescence emission and electronic one-photon absorption (UV-vis) spectra for Rh800 in DMSO (**Fig. 2b-c**) with FCclasses3,[18] testing the effects of different solvent models (**Fig. S2**), functionals (**Fig. S3**), and basis sets (**Fig. S4**). We find that calculations at the B3LYP/6-31G(d,p) level afford a good balance of accuracy and cost for electronic spectra, in good agreement with previous literature.[19, 47, 48] Importantly, we find that the choice of solvent model is vital, consistent with the Stark effect.[49, 50] Our computations show that the Solvation Model with Density (SMD),[51] which adds empirical terms for intermolecular interactions beyond a pure dielectric picture, yields an appreciably more accurate description of electronic spectra than polarizable continuum models (**Section S2**; **Fig. S2**).

As shown in **Fig 2b-c**, the calculated electronic spectra of Rh800 are dominated by the 0-0 transition ($|\langle 0'|0\rangle| = 0.73$), which is calculated to be 714 nm in DMSO (experimentally, ~711-713 nm in DMSO),[24, 52] with weaker vibronic sidebands giving rise to the overall shape of the spectra. These findings are in good agreement with previous work on chromophores with charge-transfer resonance forms.[53, 54] From Marcus-Hush theory, strong mixing (i.e., Robin-Day Class III) between the diabatic potentials of these resonance forms gives rise to the adiabatic ground- and excited-state potentials $S_0$ and $S_1$, which have zero displacement along the charge-transfer

coordinate in the totally symmetric case.[54, 55] As such, top-down theoretical models also predict spectra that are dominated by the 0-0 transition for similar chromophores, in strong agreement with our computational and experimental results.

**B. Computing BonFIRE activity**

After validating our DFT methods, we then turned to the problem of computing BonFIRE spectra. As we established in **Section II**, BonFIRE can be described as electronic one-photon absorption from a vibrationally pre-excited state. Thus, a theoretical BonFIRE spectrum can be obtained by calculating the one-photon absorption spectrum following vibrational excitation in each normal mode individually, then multiplying by the MIR absorption cross-section of each normal mode (**Eq. 5**). To make these calculations scalable towards high-throughput screening with large molecule sets, we developed a fully automated pipeline for BonFIRE spectrum calculations, which we call AutoDFT (**Fig. 3**). In AutoDFT, we leverage several validated, industry-standard computational chemistry tools (OpenBabel, Gaussian16, and FCclasses3)[18, 26, 27] and integrate them into a single Python wrapper, requiring only an input ChemDraw structure for all calculations. By making our code openly accessible, we hope to encourage broader adoption of such tools (especially by fellow experimentalists), since barriers to accurate and reliable computational methods remain a challenge even today.[54] We are also highly motivated to continue developing AutoDFT as an open, accessible tool for calculating vibrational and vibronic spectra.

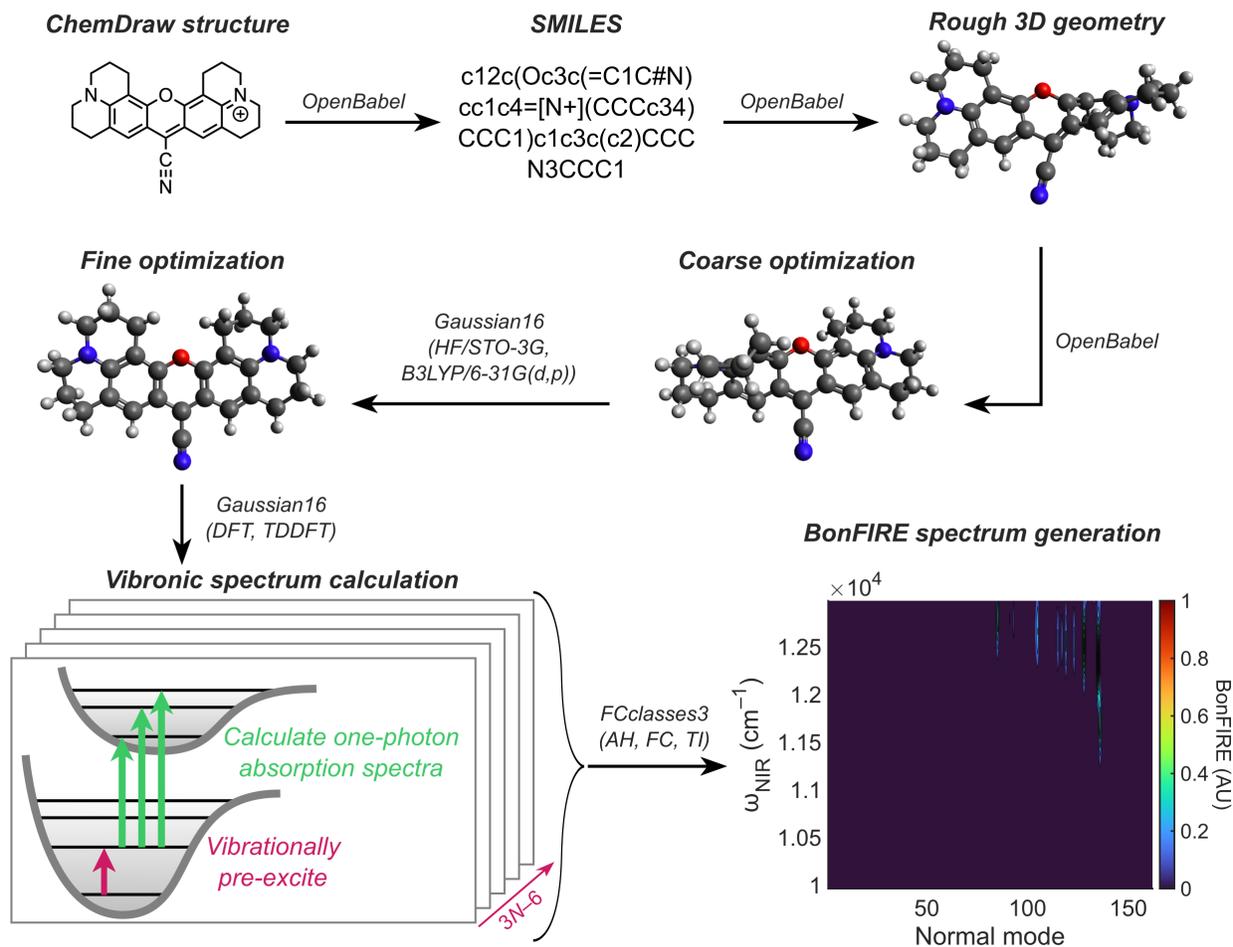

**Figure 3.** AutoDFT: a fully automated computational pipeline for batch calculations of BonFIRE vibronic spectra. AH, adiabatic Hessian; FC, Franck-Condon; HF, Hartree-Fock; SMILES, simplified molecular input line entry system; TI, time-independent.

We first applied our AutoDFT pipeline to three well-studied BonFIRE dyes: Rh800, Cy5.5, and ATTO680 (**Fig. 4**).[14, 16] To calculate BonFIRE intensity as a function of MIR frequency, we take the maximal intensity projection along the NIR axis in the electronic pre-resonance regime (e.g., $\omega_{NIR}$ < 13,000 cm$^{-1}$ for Rh800), taking the peak in the NIR spectrum as BonFIRE signal. We plot predicted spectra for each of these molecules as a function of $\omega_{MIR}$ in **Fig. 4**, comparing against our previously reported experimental spectra.[14]

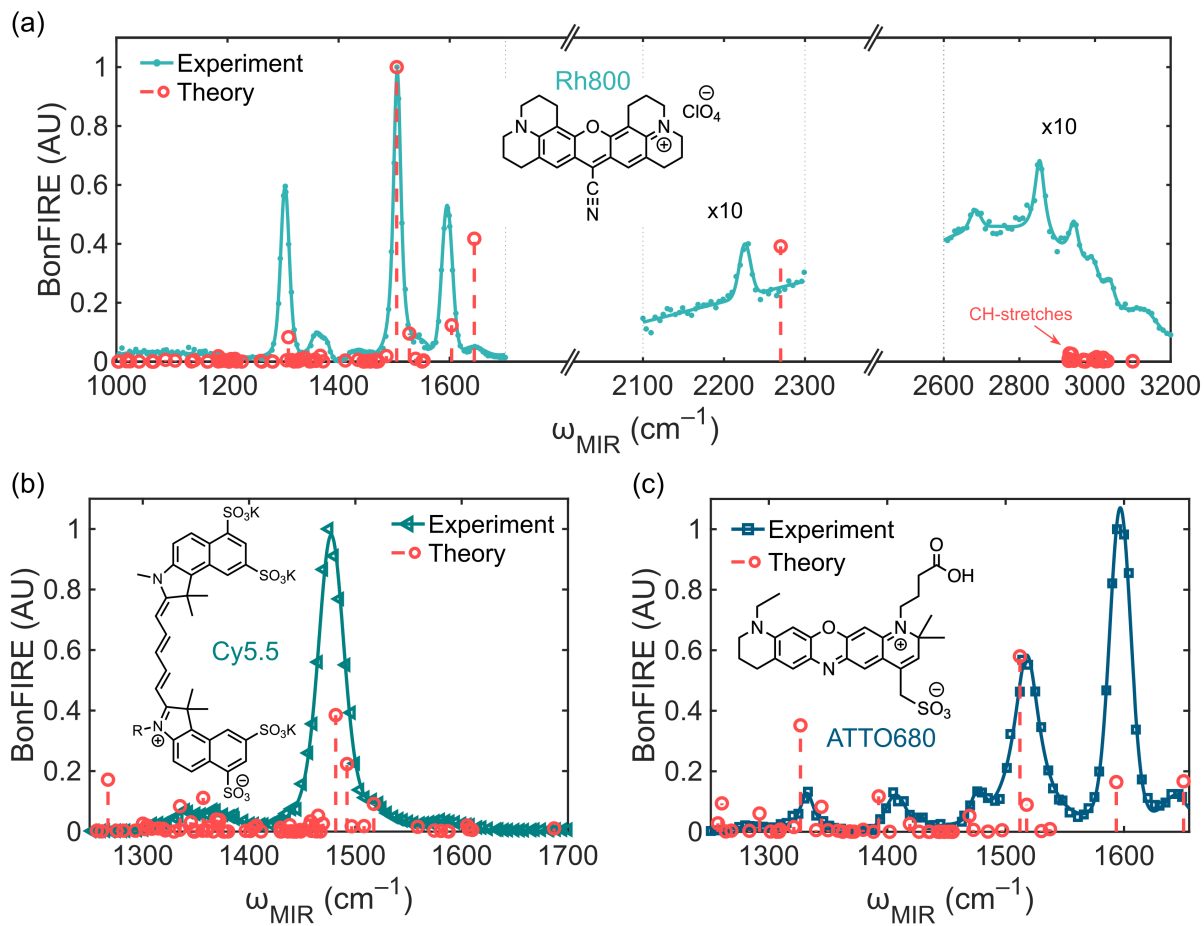

**Figure 4.** Comparison of computed and experimental BonFIRE spectra for (a) Rh800, (b) Cy5.5, and (c) ATTO680 (structures inset). Both calculated and experimental intensities are scaled up 10-fold in the cell-silent and CH-stretching regions in panel (a). Experimental spectra (normalized to 1) were obtained from a previous report.[14] Computed MIR frequencies were scaled by 0.97. Computed BonFIRE spectra were scaled in intensity to (a) 1, (b) 0.39, and (c) 0.58 to better match experimental intensities across the entire spectrum.

Across the fingerprint region (~800-1800 cm$^{-1}$), we generally observe good agreement in frequency (within ~5 cm$^{-1}$) and fair agreement in intensity (within an order of magnitude) between our computed and experimental spectra (**Fig. 4a-c**), consistent with our calculations of steady-state spectra (**Fig. 2**). Notably, there are no strongly active BonFIRE modes below ~1300 cm$^{-1}$ for Rh800, in good agreement with our experimental data at 100 μM in DMSO (**Fig. 4a**), despite several predicted MIR-active modes in the 1000-1300 cm$^{-1}$ range (**Fig. 2a**). These results are particularly illuminating, as in previous BonFIRE work with higher concentrations of Rh800 (~50 mM) in thin polymer films, we have observed a handful of relatively strong peaks can be detected in the 1000-1300 cm$^{-1}$ range;[16] we now suspect these peaks may result from dimers or other aggregates, since our present computational results (on a single molecule of Rh800) and dilute spectra in DMSO are in good agreement. Collectively, these results suggest that the 1300-1800 cm$^{-1}$ region appears to be much more promising for continued single-molecule spectroscopy with BonFIRE and related methods due to the large FCFs.[8, 13, 16, 25]

Looking beyond the fingerprint region, we note that the slanted baseline in the cell-silent region (1800-2300 cm$^{-1}$; **Fig. 4a**, middle) and broad background in the CH-stretching region (2600-3200 cm$^{-1}$; **Fig. 4a**, right) have been previously assigned to non-degenerate resonance-enhanced two-photon absorption (NDR-TPA),[14, 56] which is not currently accounted for in our current computational methods. Moreover, highly accurate calculations of nitrile-stretching frequencies remain a challenge for DFT calculations within the harmonic approximation (due to the nitrile's highly anharmonic nature),[14, 15] as evidenced by the ~40 cm$^{-1}$ gap between the experimental and calculated nitrile frequencies even after applying a scaling factor of 0.97 (**Fig. 4a**, middle); empirically, we find that a scaling factor of ~0.953 yields better agreement for nitriles in particular. However, the agreement in intensity between experiment and theory for the nitrile stretch of Rh800 is good (within a factor of two, after accounting for NDR-TPA), suggesting that the computed IR activities and FCFs are still accurate. We note that our current work employs an adiabatic Hessian model,[18] treating the FCFs as functions of the dimensionless displacement, mode distortion, and the Duschinsky rotation, which appears to be valuable in calculating BonFIRE spectra.

**C. Violations of the resonance condition originate from combination modes**

Notably, the computed spectrum of Rh800 differs significantly from the experimental data in the CH-stretching region (2600-3200 cm$^{-1}$; **Fig. 4a**, right), in both $\omega_{MIR}$ and intensity. Quite clearly, pre-excitation in the CH-stretching fundamental modes is not predicted to produce appreciable BonFIRE signal, due to the small FCFs of the CH-stretches.[57] However, several prominent peaks were observed experimentally (**Fig. 4a**, right).[14]

To further investigate this disparity, we examined the $\omega_{NIR}$-dependences of our experimental and computed spectra (**Fig. 5**).[14] In the $\omega_{NIR}$-domain, we quantify the resonance condition by the position of the peak in the BonFIRE spectrum ($\omega_{NIR}^{peak}$) relative to the UV-vis absorption maximum ($\omega_{UVV}$). When the resonance condition is maintained, the energy difference between the BonFIRE peak and the UV-vis absorption maximum ($\Delta\omega = \omega_{UVV} - \omega_{NIR}^{peak}$; numbers above horizontal double-arrows in **Fig. 5**) matches the MIR photon energy ($\omega_{MIR}$; labels in the top-right legends in **Fig. 5**). Experimentally, for modes in the fingerprint and cell-silent regions, the resonance condition is consistently maintained (i.e., $\Delta\omega = \omega_{MIR}$; **Fig. 5a**). In sharp contrast, the resonance condition is visibly broken for modes in the 2600-3200 cm$^{-1}$ region,[14] where experimental BonFIRE $\omega_{NIR}$ spectra show a peak redshifted by only $\Delta\omega \approx 1500$ cm$^{-1}$ from the UV-vis absorption maximum (**Fig. 5b**), which we had termed 'VRCs' in **Section I** ($\Delta\omega \neq \omega_{MIR}$; see **Fig. 1d**).

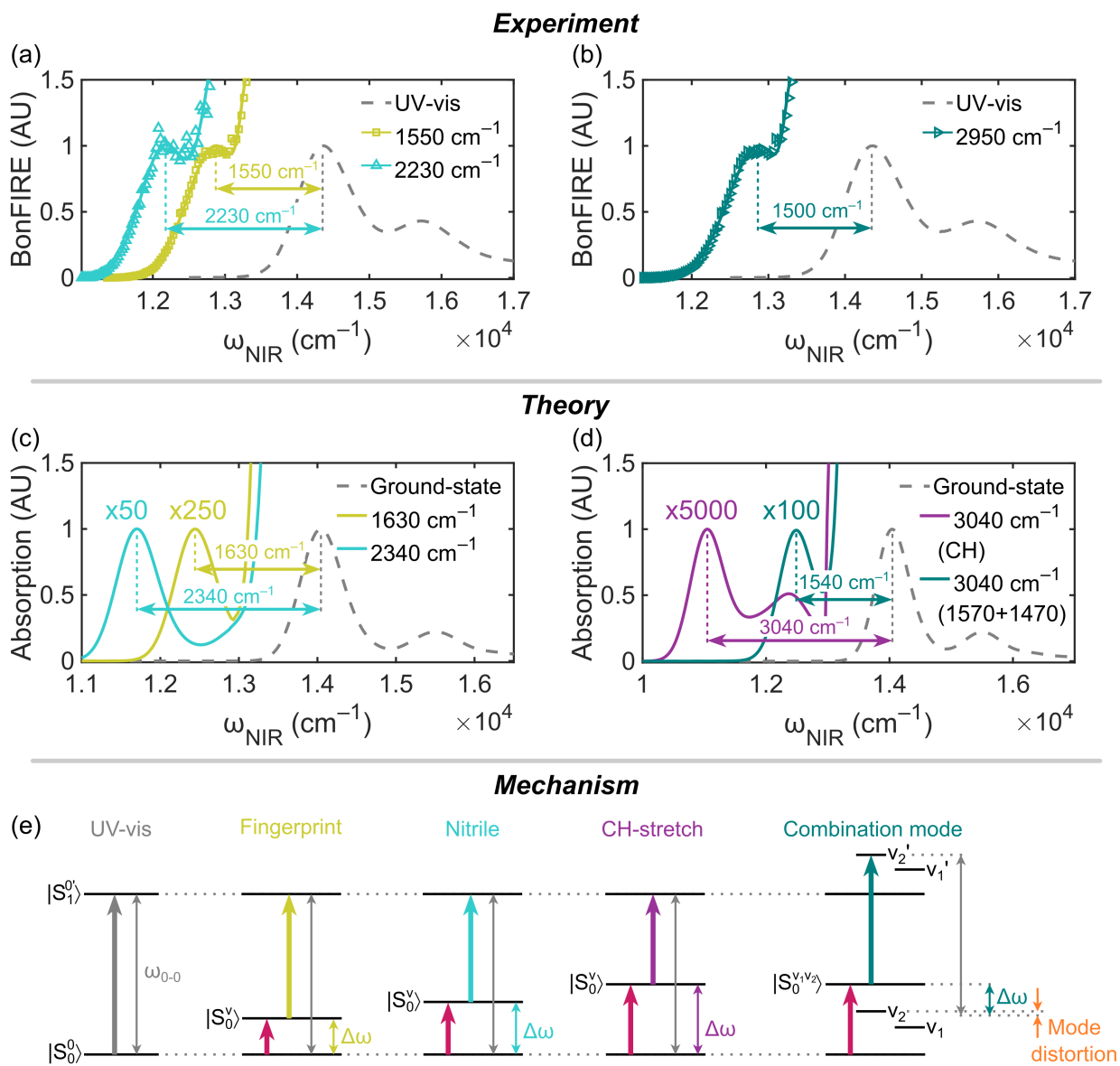

**Figure 5.** Mechanism of VRCs. (a-b) Experimental BonFIRE $\omega_{NIR}$ spectra of Rh800 at (a) $\omega_{MIR}$ = 1550 cm$^{-1}$ (yellow), 2230 cm$^{-1}$ (cyan), and (b) 2950 cm$^{-1}$ (teal).[14] (c-d) Computed BonFIRE $\omega_{NIR}$ spectra of Rh800 following vibrational pre-excitation in (c) ring-breathing (yellow) and nitrile-stretching (cyan) modes and (d) CH-stretching (purple) and ring combination (teal) modes, as compared to spectra computed from the ground state (gray). The horizontal double-sided arrows denote the energy differences between the peaks in BonFIRE and UV-vis. (Computed spectra were normalized to the peak in the pre-resonance regime; unscaled spectra are shown for reference in **Fig. S5**.) (e) Energy-level diagrams for BonFIRE. $\Delta\omega$ denotes the energy gap between the $\omega_{NIR}$ peaks in UV-vis and BonFIRE, as marked

in panels (a-d). The VRCs are consistent with up-conversion from a combination mode, where only one of the vibrational quanta changes (the other remains excited in $S_1$).

We then examined our computed $\omega_{NIR}$ spectra, comparing electronic one-photon absorption spectra without pre-excitation (analogous to UV-vis) to one-photon electronic absorption spectra from vibrationally pre-excited states (analogous to BonFIRE; **Fig. 5c-d**). We note that these spectra are calculated directly with FCclasses3, meaning that the vibrational frequencies are all unscaled (and thus are ~3-5% higher) relative to experimental spectra. These computed spectra are also scaled in intensity to permit visualization of the shape of the spectrum in the pre-resonance regime, with scaling factors inset (**Fig. 5c-d**) and unscaled spectra shown in **Fig. S5** (**Section S3**).

Examining the computed $\omega_{NIR}$ spectra for a representative fingerprint mode (1630 cm$^{-1}$) and the nitrile stretch (2340 cm$^{-1}$), we indeed observe peaks in the electronic pre-resonance regime (**Fig. 5c**, yellow and cyan curves), indicating that these modes are appreciably coupled. Moreover, these peaks are redshifted from the peak in the ground-state absorption spectrum by exactly the energy of the excited vibrational mode ($\Delta\omega = \omega_{MIR}$), in excellent agreement with our experimental results (**Fig. 5a**).

We then examined the computed $\omega_{NIR}$ spectrum following pre-excitation in a CH-stretching mode (3040 cm$^{-1}$). CH-stretching modes are generally thought to have weak vibronic coupling,[57] and our computations confirm this quite readily, as evidenced by the 5000-fold scaling needed to see the peak in the pre-resonance regime (**Fig. 5d**, purple curve). Furthermore, as with the other fundamental modes, pre-excitation in a CH-stretch is calculated to yield a redshifted peak matching the vibrational mode energy ($\Delta\omega = \omega_{MIR}$), indicating that no VRC is expected for a CH-stretching pre-excitation. Thus, in $\omega_{MIR}$, $\omega_{NIR}$, and overall intensity (**Figs. 4a**, **5b**, and **5d**), the peaks that we observe experimentally in the 2600-3200 cm$^{-1}$ region are not consistent with pre-excitation in CH-stretching modes.

However, when we pre-excite in a combination mode ($|v_{comb}\rangle = |v_1 v_2\rangle$), also with energy totaling 3040 cm$^{-1}$ (comprising pre-excitation in modes with energies of $\omega_1 = 1470$ cm$^{-1}$ and $\omega_2 = 1570$ cm$^{-1}$), we observe a peak redshifted by only 1540 cm$^{-1}$, and no peak is observed with a redshift of 3040 cm$^{-1}$ (**Fig. 5d**, teal curve). Moreover, this redshifted peak is much stronger than that of the CH-stretch, with a vibronic coupling strength comparable to a fingerprint mode (evidenced by the 100-fold scaling; **Fig. 5d**). To our knowledge, this is the first computational reproduction of a VRC, and it is in excellent agreement with our experimental spectra (**Fig. 5b**), strongly supporting the assignment of the peaks that we observe in the 2600-3200 cm$^{-1}$ region as combination modes.

## D. Franck-Condon factors and resonance conditions of combination modes

Though perhaps counterintuitive on first glance, these calculations are highly informative. First, considering the fingerprint pre-excitation, the dominant electronic transition is calculated to be $|S_0^v\rangle \to |S_1^{0'}\rangle$, which is responsible for the BonFIRE peak at $\omega_{NIR}^{peak} = \omega_{0-0} - \omega_{MIR}$ (**Fig. 5e**, yellow), obeying the resonance condition. A subtle but crucial detail of this transition is that, as the electronic excitation occurs ($|S_0\rangle \to |S_1\rangle$), the molecule transitions from a vibrationally excited state to the vibrational ground state ($|v\rangle \to |0'\rangle$), indicating that formally, the molecule loses vibrational energy during the electronic absorption; the 'lost' vibrational energy is converted into electronic energy to allow for the transition to $|S_1^{0'}\rangle$ (this is distinct from spontaneous vibrational relaxation). This conversion of MIR-excited vibrational energy into electronic energy is intrinsic to BonFIRE, and it is common to the other fundamental band excitations (nitrile and CH-stretch; **Fig. 5e**, cyan and purple).

However, in the combination mode case, no peak is calculated at $\omega_{0-0} - \omega_{MIR}$ (**Fig. 5d**, teal, curve), implying that the $|S_0^{v_1 v_2}\rangle \rightarrow |S_1^{0'}\rangle$ transition (converting the energy of both modes into electronic energy) is not favored. To understand why, we recall (as we discussed in **Section IV.A**) that the electronic absorption spectra of large organic dyes tend to be dominated by the 0-0 transition and consequently have small displacements ($\delta$) along the normal coordinates, meaning that the ground-state and excited-state geometries are similar.[53] FCFs are primarily functions of the displacement, and as reported by Tokmakoff and co-workers, they can be reasonably approximated as purely functions of the displacement (e.g., the Huang-Rhys factor, $S_v = \delta_v^2/2$).[19,24] Indeed, for the transition $|S_0^{v_1 v_2}\rangle \rightarrow |S_1^{0'}\rangle$, starting from a pre-excited combination mode and describing conversion of both modes into electronic energy (i.e., upholding the resonance condition), the combination of two small displacements leads to a vanishingly small calculated FCF ($|\langle 0'|v_1 v_2\rangle| \approx 0$).

The question then becomes: what transition(s) do we observe? When we examined the other FCFs in the combination mode pre-excitation, we saw that the FCF for the $|S_0^{v_1 v_2}\rangle \rightarrow |S_1^{v_2'}\rangle$ transition (converting only one mode of the combination; $|\langle v_2'|v_1 v_2\rangle| = 0.062$) is comparable to the FCF of the $|S_0^{v_1}\rangle \rightarrow |S_1^{0'}\rangle$ transition from the equivalent fingerprint mode ($|\langle 0'|v_1\rangle| = 0.078$), being governed by the displacement along the normal coordinate of $v_1$. In contrast, the FCF for the $|S_0^{v_1 v_2}\rangle \rightarrow |S_1^{v_1'}\rangle$ transition (conversion of $v_2$) is much smaller ($|\langle v_1'|v_1 v_2\rangle| = 0.0195$), yielding only a minor (~10%) contribution to the overall up-conversion spectrum. Indeed, it is the $|S_0^{v_1 v_2}\rangle \rightarrow$

$\left|S_1^{v_2'}\right\rangle$ transition (conversion of $v_1$) that dominates the computed BonFIRE spectrum in the combination mode pre-excitation case, giving rise to a VRC (**Fig. 5e**, teal).

To further confirm this reasoning, we pre-excited in the first overtone ($v = 2$) of the nitrile stretch (with fundamental FCF $|\langle 0'|v_{CN}\rangle| = 0.12$) and again computed the electronic one-photon absorption spectrum (**Fig. S6**). Consistent with the combination mode case, a much smaller FCF is calculated for converting both quanta of the nitrile stretch into electronic energy ($|\langle 0'|2v_{CN}\rangle| = 0.024$) than for converting only one quantum of the nitrile stretch ($|\langle v_{CN}'|2v_{CN}\rangle| = 0.16$). This overtone calculation demonstrates that the conversion of one quantum of vibrational energy is nearly two orders of magnitude greater in transition intensity than the conversion of both, even when both quanta are in a strongly coupled mode (**Fig. S6**). As such, it appears to be generally favored to convert only one vibrational quantum into electronic energy upon electronic excitation, meaning that VRCs should be expected for BonFIRE with anharmonic transitions.

Lastly, we explored the energetic basis of the calculated 1540 cm$^{-1}$ redshift for the $\left|S_0^{v_1 v_2}\right\rangle \rightarrow \left|S_1^{v_2'}\right\rangle$ transition (**Fig. 5d**, teal curve). This redshift was surprising, given that the combination mode (total energy 3040 cm$^{-1}$) comprises modes with energies of $\omega_1 = 1470$ cm$^{-1}$ and $\omega_2 = 1570$ cm$^{-1}$. Based on the above FCF analysis, it is strongly favored to convert only mode $v_1$ into electronic energy. In this case, one might reasonably expect a redshift of $\omega_1 = 1470$ cm$^{-1}$ instead of our calculated value of 1540 cm$^{-1}$. However, upon close inspection of our TDDFT calculations, the remaining difference of 70 cm$^{-1}$ is caused by the difference in the energy of mode $v_2$ between $S_0$ and $S_1$ (i.e., $\omega_2 = 1570$ cm$^{-1}$, but $\omega_2' = 1500$ cm$^{-1}$). This energy difference results from a change in the shape of the potential for $v_2$ between electronic states (generally, a broadening of the potential or reduction of the vibrational frequency in $S_1$), which is known as the "mode

distortion" (**Fig. 5e**, orange).[58] Therefore, even though mode $v_2$ remains excited and is not converted into electronic energy, its 70 cm$^{-1}$ mode distortion influences the peak position in the calculated BonFIRE spectrum (**Fig. 5d**, teal curve), thus leading to a calculated redshift of 1540 cm$^{-1}$ from a 3040 cm$^{-1}$ (1570 + 1470) combination mode. Applying a scaling factor of 0.97 to the vibrational frequencies, the energy of the pre-excited combination mode is ~2950 cm$^{-1}$, and the calculated redshift is ~1500 cm$^{-1}$, in excellent agreement with our experimental results (**Fig. 5b**).

Integrating these findings, we derive an analytical expression for the expected NIR peak position ($\omega_{NIR}^{peak}$) for a combination mode in BonFIRE as:

$$\omega_{NIR}^{peak} = \omega_{UVV} - [\omega_1 + (\omega_2 - \omega_2')] \tag{6}$$

In **Eq. 6**, mode $v_1$ is the converted mode and mode $v_2$ is the 'spectator' mode, whose mode distortion contributes to the BonFIRE peak position (**Fig. 5e**). Accordingly, the term in square brackets in **Eq. 6** describes the expected redshift from the UV-vis peak ($\Delta\omega$), given by $\Delta\omega = \omega_{UVV} - \omega_{NIR}^{peak}$. **Eq. 6** can also be rearranged as $\omega_{NIR}^{peak} = \omega_{UVV} - \omega_{MIR} + \omega_2'$, since $\omega_1 + \omega_2 = \omega_{MIR}$ in order for the combination mode to be MIR-resonant. If the energies of the involved modes are precisely known (e.g., if measured by a complementary method[59]), it is also possible for BonFIRE to experimentally inform the mode distortion of the non-converted mode.

Collectively, our calculations provide new insights into the FCFs and resonance conditions of vibronic transitions involving MIR-excited combination states. We find that the conversion of one vibrational quantum into electronic energy is strongly favored over multiple conversions, providing a mechanistic basis for VRCs. Our spectral calculations further establish the involvement of mode distortion in determining $\omega_{NIR}^{peak}$ in BonFIRE (**Eq. 6**), illuminating the complexity of the resonance conditions for anharmonic transitions.

**V. POTENTIAL APPLICATIONS**

As shown in **Section IV**, computations can yield deep, quantitative insights into experimental spectra and serve as invaluable aids for interpreting complex results. Beyond detailed analysis, computations can also be performed at scale and predictively, and we envision several possible uses of such tools towards the design and screening of new molecular probes. We now highlight a few such potential applications.

**A. Design and screening of local environment sensors**

A well-established application of vibrational spectroscopic probes is the quantitative reporting of the characteristics of local environments, such as electric fields,[1, 15] hydrogen bonding,[36, 60, 61] temperature,[52, 62] and pH.[63] The design of these probes is generally rational, beginning from established physical principles like the vibrational Stark effect, which has proven effective but can be time-consuming and labor-intensive. We envision a complementary approach for the discovery of new vibrational sensors, leveraging computational tools for high-throughput screening (**Fig. 6a**). Of course, the success of such an approach is heavily reliant on accurate computations, but with recent advancements in calculating vibrational spectra with DFT,[60] molecular dynamics,[64] and machine learning,[65] we believe computational methods hold significant promise for identifying environment-sensitive molecular vibrations. Coupled with AutoDFT, these methods should be extensible to predictions of BonFIRE local environment sensors, which could be used for fundamental studies of chemical reactions (for example, at interfaces or on nanoparticles)[66, 67] or applied towards mapping biological activity (such as enzyme kinetics or membrane protein dynamics).[68]

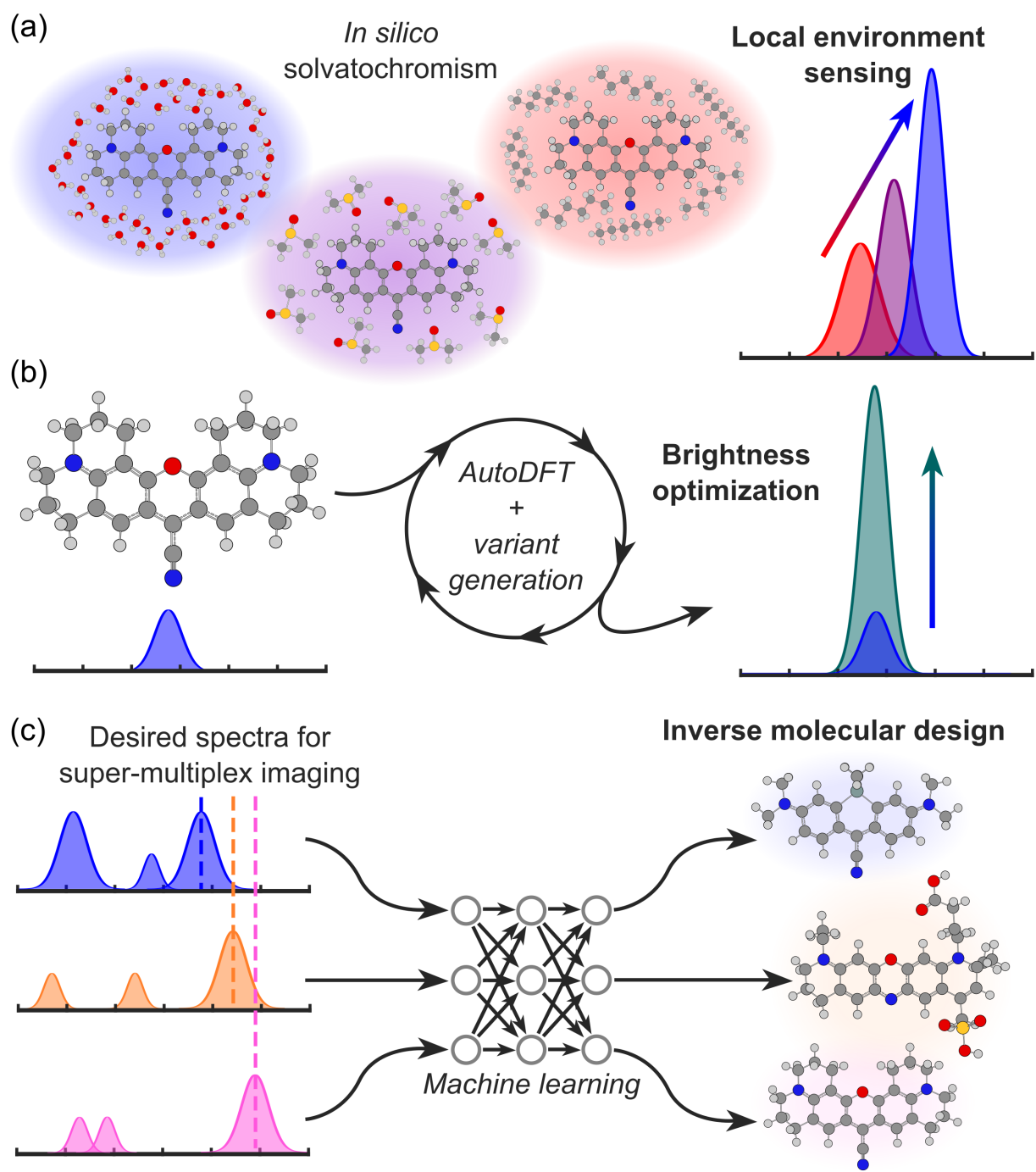

**Figure 6.** Potential applications of computational predictions of BonFIRE spectra. (a) *In silico* solvatochromism as a means of discovering vibrational local environment sensors. (b) Iterative brightness optimization via AutoDFT and chemical variant generation. (c) Inverse molecular design for super-multiplex imaging, using machine learning to predict target molecular structures from desired spectra.

**B. Iterative optimization of brighter BonFIRE probes**

Our primary usage of our AutoDFT pipeline in this work has been in describing the relative intensities of vibrational modes, towards calculating BonFIRE spectra (**Fig. 4**). However, one can also envision using this pipeline to quantify absolute intensities, as part of an iterative refinement process towards developing brighter BonFIRE probes. In a scheme similar to directed evolution,[69] the combination of AutoDFT with a tool for *de novo* chemical structure variant generation[70, 71] would allow for rapid screening of various chemical motifs and their effects on BonFIRE brightness (**Fig. 6b**). Such an approach could not only grant brightness improvements towards broader single-molecule spectroscopy,[16] but also provide fundamental insights into how discrete chemical modifications can fine-tune BonFIRE signal,[19, 72] which is largely underexplored at the time of writing.

**C. Inverse molecular design for super-multiplex imaging**

Another well-established application of vibrational probes is super-multiplex imaging, where narrow vibrational bands facilitate the imaging of many different probes in a single round of labeling.[14, 72, 73] Naturally, it is desirable in such an application to have a palette of probes with well-separated vibrational peaks, which is greatly assisted by the ability to fine-tune vibrational frequencies at will. For nitrile vibrational probes, the effects of structural modifications on frequency have been well-explored.[72, 74] However, for molecular vibrations in the fingerprint region, the effects of structural tuning on vibrational frequencies are less well-explored.

In this context, we see the potential for inverse molecular design,[75] where machine learning can serve as a bridge between desired spectra (with distinct, well-separated peaks) and potential molecular structures that give rise to the desired spectra.[76] Of course, the training and validation of such a model will require extensive experimental data; we see a clear role for an open BonFIRE spectral database in achieving this goal, and we are currently acquiring and tabulating reference spectra to facilitate such an application. This type of model could also offer new insights into how molecular structure informs vibronic spectra, providing a chemical basis for BonFIRE as a fingerprinting tool. Additionally, though vibrational imaging has already been shown to reach the 10-20-color level,[14, 73] an inverse design approach could allow for a significant palette expansion, possibly approaching 100 colors in a single shot. In particular, the high dimensionality of BonFIRE (allowing for differentiation in $\omega_{MIR}$, $\omega_{NIR}$, vibrational lifetime, and emission spectrum)[14] appears particularly promising towards this goal.

## VI. CONCLUSION

Computational methods are indispensable tools to aid in the interpretation of complex spectra and act as guides for future experiments. In this Perspective, we briefly reviewed the theory of bond-selective fluorescence (BonFIRE) spectroscopy. We then developed an automated computational pipeline to enable predictions of BonFIRE spectra, yielding decent agreement with experimental spectra. Our computations revealed that VRCs in the 2600-3200 cm$^{-1}$ region are consistent with the MIR excitation of combination modes, showing that the conversion of two vibrational quanta into electronic energy upon electronic excitation is substantially disfavored relative to the conversion of a single quantum for large dye molecules. Finally, we highlighted a few potential applications for automated predictions of BonFIRE spectra towards quantitative local environment sensing, iterative brightness optimization, and inverse molecular design for super-multiplex imaging.

**SUPPLEMENTARY MATERIAL**

See the supplementary material for detailed methods, validation of DFT methods, supplemental figures, and references.

**ACKNOWLEDGEMENTS**

P.A.K. thanks Dr. Tomislav Begušić, Dr. Xuecheng Tao, Dr. Chenghan Li, Dr. Nathanael Kazmierczak, Jax Dallas, and Dr. Jacob Kirsh for fruitful conversations on broad-ranging topics in theoretical chemistry. P.A.K. is grateful for financial support from a National Science Foundation (NSF) Graduate Research Fellowship (DGE-1745301) and a Hertz Fellowship. R.E.L. is grateful for funding provided by the Arnold and Mabel Beckman Foundation, 2025 Arnold O. Beckman Postdoctoral Fellowship in Chemical Instrumentation. The computations presented here were conducted with the Resnick High Performance Computing Center, a facility supported by the Resnick Sustainability Institute at the California Institute of Technology. We thank the Caltech Beckman Institute Laser Resource Center for research resources. This work was supported by a National Institutes of Health Director's New Innovator Award (DP2 GM140919-01 for L.W.), an Alfred P. Sloan Research Fellowship (L.W.), and an NSF CAREER Award (CHE-2240092 for L.W.). L.W. is a Heritage Principal Investigator supported by the Heritage Medical Research Institute at Caltech.

**AUTHOR INFORMATION**

**Conflict of Interest**

The authors have no conflicts to disclose.

**Author Contributions**

Conceptualization and writing – original draft: P.A.K. and L.W. (equal). Methodology, formal analysis, and investigation: P.A.K. (lead); all authors (supporting). Validation and writing – review & editing: all authors. Funding acquisition, project administration, resources, and supervision: L.W.

**DATA AVAILABILITY**

The source code for AutoDFT (https://github.com/pkocheril/AutoDFT/) and our MATLAB data processing code (https://github.com/pkocheril/BonFIRE-Processing-Code) are openly available on GitHub. The data that support the findings of this study are available from the corresponding author upon reasonable request.